\documentclass[twocolumn,showpacs,preprintnumbers,amsmath,amssymb,showkeys]{revtex4}

\usepackage{graphicx}% Include figure files
\usepackage{dcolumn}% Align table columns on decimal point
\usepackage{bm}% bold math
\usepackage{bezier}
\usepackage{slashed}

\newcommand{\expectation}[1]{\langle#1\rangle}

\newcommand{\expectationfit}[1]{\left\langle#1\right\rangle}

\newcommand{\Or}{\mathrm{O}}
\newcommand{\rmd}{\mathrm{d}}

\begin{document}

\title{Calculation of the electron two-slit experiment using a quantum mechanical variational principle}
\author{Alan K. Harrison}
\affiliation{Los Alamos National Laboratory,
Los Alamos, New Mexico   87545}

\date{\today}% It is always \today, today,

\begin{abstract}
A nonlocal relativistic variational principle (VP) has recently been proposed as an alternative to the Dirac wave equation of
standard quantum mechanics. 
We apply that principle to the electron two-slit experiment. 
The detection system is modelled as a screen made of atoms, any one of which can be excited by the incident electron,
but we avoid restricting the detection mechanism further.
The VP is shown to predict that, at the time the electron reaches the screen,
its wavefunction will be localized to the neighborhood of a single atom, resulting in a position-type measurement.
In an ensemble of such experiments
(``identically prepared'' except that the initial phase of the wavefunction---the hidden variable in the VP formulation---
is sampled over the expected uniform distribution),
the distribution of measured positions will reproduce the interference pattern predicted by the Dirac equation.
We also demonstrate that with a detection system designed fundamentally to detect the electron's transverse wavelength
rather than its position, the VP predicts that one such mode will be detected, that is, a wavelength measurement will result.
Finally, it is shown that these results are unchanged in the ``delayed choice'' variant of the experiment.
\end{abstract}
\maketitle

\section{Introduction}
An alternative version of quantum mechanics has recently been proposed \cite{Harrison_2012a} in the form of a nonlocal relativistic variational principle (VP).
The VP is intended to replace both the wave equation and the measurement-induced collapse process of standard quantum mechanics;
that is, it is proposed as a unified theory valid regardless of whether a measurement is being made.
To test whether it can adequately perform both roles, we will apply it to predict the outcome of the electron two-slit diffraction experiment.\cite{Mollenstedt_Jonsson}

In the original form of that experiment,
%with a beam of electrons incident on a barrier containing two parallel slits,
the detection screen
shows the diffraction/interference pattern predicted by wave mechanics.
If the experiment is conducted with a low beam intensity, so that the electron arrival positions
at the screen can be observed, the position data sum up
to the same interference pattern that the higher-intensity beam produces, \cite{Merli_Missiroli_Pozzi}
even when the experiment is conducted with so low a beam intensity that the electrons pass the barrier one at a time. \cite{Tonomura}
%This is a challenge to explain in terms of wave-or-particle complementarity, because the observed collision positions make sense in terms of the electron as a particle, but the interference pattern is a manifestation of its wave character. 
%Put another way, the interference pattern implies that each electron went through both slits at once, but the position measurements shows that when it reached the screen, it had a single well-localized position.

Standard quantum mechanics (SQM) explains the position measurement as a result of the collapse process
and the interference pattern as a result of the Born rule that it obeys,
but fails to give an intuitively satisfying answer to detailed questions about the electron's path.
\cite{Feynman,Wheeler_1979}
We will show that the VP predicts the results of the experiment as well, but in a unified way, and that it gives greater insight into such questions.

The VP asserts that Nature minimizes the sum of two spacetime integrals $A_1$ and $\epsilon A_2$, both of which depend on the
wavefunction of the system under study, possibly including an entangled measurement apparatus. 
The first term measures
the deviation of the solution from compliance with the SQM wave equation (the Dirac equation, since the theory is currently limited to fermions). $A_2$ is a measure of the position-momentum (or time-energy) uncertainty of the wavefunction, which we expect to be larger for a superposition of eigenstates of the operator corresponding to the measurement than for a single mode.
The result of the optimization process is that an undisturbed system satisfying the Dirac equation will continue to do so,
but a change in the external fields it experiences---as in an experiment---may cause it to make a transition from one state to another.
%$A_1$ also has the effects of limiting the rate of transitions (by penalizing rapid changes),
%and (approximately) enforcing the Born rule. 

According to this view, the electron may be in a superposition of position eigenstates when it passes through the slits, but the tendency to
minimize uncertainty causes the wavefunction to decay to a single such state by the time it reaches the detection screen.
This paper tests that prediction by performing an approximate VP calculation.

\section{The variational principle}
To define the VP compactly, we will define some mathematical structures.
For an $N$-particle system with wavefunction $\psi$
and an operator $\mathcal{O}$ that depends on $K$ different spacetime coordinates for each particle,
we define the $K$-position expectation integral
\begin{multline}
\label{define_I_K}
I_K(\mathcal{O}) \equiv
\int \left( \prod_{n=1}^N \, \prod_{k=1}^K \rmd^4 x_{nk} \right) \,\,
\left( \prod_{k=1}^K \psi^\dagger(x_{1k}, \ldots x_{Nk}) \right)
\\
\mathcal{O}(x_{11}, x_{12}, \ldots, x_{1K}, x_{21}, \ldots, x_{NK})
\left( \prod_{k=1}^K \psi(x_{1k}, \ldots x_{Nk}) \right)
\\
\prod_{n=1}^N \, f_K(x_{n1}, x_{n2}, \ldots x_{nK} )
\, .
\end{multline}
Here $f_K$ is unity if $K=1$; otherwise it 
enforces the spacelike separation of all $K$ points for a given particle $n$:
\begin{multline*}
%\label{final_f_4pt}
f_K( \{x_k\} )
\equiv
\frac
{\prod^{K-1}_{k=1} \prod^K_{l=k+1} u \left[- \left(x_k^\mu-x_l^\mu\right) \left(x_{k\mu}-x_{l\mu}\right) \right]}
{W_K( \{x_k-x_l : 1\! \le \! k \! < \! l \! \le \! K\} )}
\, ,
\end{multline*}
where
we use the summation convention for repeated Greek indices, which run from 0 to 3;
$u(z)$ is the unit step (Heaviside) function;
and the weight function $W_K$ is chosen so that $f_K$ has the property
\begin{equation}
\label{integrate_final_f_4pt}
\int \rmd x_2^0 \int \rmd x_3^0 \ldots \int \rmd x_K^0 \,\,
f_K( \{x_{nk}\} )
= 1
\, .
\end{equation}
(For any value of $K$, $W_K$ is expected to be a universal function, but its precise form is unknown for $K>2$;
see discussion in \cite{Harrison_2012a}.)
Then we define the $K$-position expectation of $\mathcal{O}$ as
\begin{equation}
\label{define_<O>_K}
\expectationfit{\expectationfit{\mathcal{O}}}_K \,
\equiv
\frac{
I_K(\mathcal{O})
}
{
I_K(1)
} \, .
\end{equation}

Now the variational principle takes the form \cite{Harrison_2012a}
\begin{equation}
\label{variational_principle}
\delta (A_1 + \epsilon A_2) = 0
\, ,
\end{equation}
in which for an $N$-particle system the first term is
\begin{equation}
\label{define_A_1_general}
A_1  =
\sum_{n=1}^N
\expectationfit{\expectationfit{\mathcal{D}_n^\dagger \mathcal{D}_n}}_1
\, ,
\end{equation}
where $\mathcal{D}_n$ is the Dirac operator $\slashed\pi/m - 1$ applied to particle $n$,
and
the integral $A_2$ in the second term is the expectation
\begin{equation}
\label{define_A_2_general}
A_2  =
\Bigg\langle \Bigg\langle
\sum_{n=1}^N
\left( \delta x^2 \delta p^2 \right)_n
\Bigg\rangle \Bigg\rangle_4 \,
\, ,
\end{equation}
of the four-point relativistic position-momentum operator
\begin{equation*}
%\label{define_delta_x_2_delta_p_2}
\delta x^2 \delta p^2
\equiv
\left\{ (x_1^\mu - x_2^\mu) \, [p_{3 \mu}(x_3) - p_{4 \mu}(x_4)] \right\}^2
\end{equation*}
applied to each particle $n$.
\section{The electron two-slit experiment}
To calculate the experiment, we suppose that
%\cite{Burke}
a single electron (particle $n=1$) is launched or first observed at time $t_i$, 
passes at time $t_a$ through slits in a planar barrier at $x^1=0$, 
and is then intercepted at $t_b$ by a detector screen in the plane $x^1=X$. 
The experiment ends at some moment $t_f > t_b$ by which time the collision point on the screen can be identified. 
The screen is composed of atoms, any of which can be excited by being hit by the electron, but the electron does not have enough energy to excite more than one atom. 
(We are intentionally vague about the nature of the ``excitation'' of an atom; our point is that in order to register the arrival of the projectile electron at a point on the screen, some atomic process must take place there. Our analysis will not depend on what that process is.) 
There are enough atoms in the screen, and their cross section is sufficient, that the probability of the electron exciting an atom is unity. 
(For simplicity, we treat the atoms as distinguishable.)

Presumably the experiment is sufficiently well isolated from the rest of the universe that we may solve the optimization problem
\eqref{variational_principle} by limiting the integrals to a domain
$\mathcal{R}$ delimited in time by $t_i$ and $t_f$, and in space by the size of the experimental setup.
%[The essential spatial requirement is that at any time, spatial integrals in \eqref{define_I_K} have the same value whether the domain of integration is $\mathcal{R}$ or infinite.]
For later use, we will also define nonoverlapping spacetime regions $\mathcal{R}_1$, $\mathcal{R}_2$, and $\mathcal{R}_3$ corresponding to the intervals
$[t_i,t_a]$, $[t_a,t_b]$, and $[t_b,t_f]$, so that $\mathcal{R} = \mathcal{R}_1 \cup \mathcal{R}_2 \cup \mathcal{R}_3$.

Consider first the prediction of the SQM wave equation (the Dirac equation) for the electronic wavefunction---that is, the prediction of SQM in the absence of any measurement process to cause collapse. That solution $\xi_{SQM}$ shows the wavefunction originating in the vicinity of the source location at $t_i$, propagating to the vicinity of both slits at $t_a$, and then spreading out into a two-slit interference pattern as it propagates into the half-space $x^1>0$. In the plane $x^1=X$, $\lvert \xi_{SQM} \rvert^2$ is exactly proportional to the diffraction pattern that is observed at the screen in a real experiment, in the limit as the number of electrons detected goes to infinity. 
Of course, $\xi_{SQM}$ does not describe an experiment with a single electron, because it will be detected at a well-defined position on the screen,
reflecting the failure of the SQM wave equation without a collapse model to describe nature. 
%As we intend to describe the single-electron experiment, it will be convenient for us to approximate the electron wavefunction $\xi$ as $\xi_{SQM}$ in the half-space $x<0$, a linear combination of states that are well-localized at $x=X$
As we intend to describe the single-electron experiment, it will be convenient for us to approximate the electron wavefunction 
%in the half-space $x>0$ 
as a linear combination of states $\xi_n$ that at time $t_a$ are localized near the slits and satisfy boundary conditions consistent with the presence of the barrier and slits, and at time $t_b$ are appreciable only in the vicinity of atom $n \, (n=2,3,4, \ldots , N)$. 
($\xi_n$ could be constructed by Green's function methods, in effect time-reversing the state that describes an electron launched from the location of atom $n$ and aimed for the slits.)
The time $C_n(t)$ in that expansion must satisfy the initial condition
\begin{equation}
\label{decompose_xi_SQM}
\xi_{SQM}(t_i, \vec x) =
\sum^N_{n=2} C_n(t_i) \, \xi_n(t_i, \vec x) \,  ,
\end{equation}
which is to say that the initial values of the coefficients $\{C_n\}$ are those that describe the interference-pattern solution of the SQM wave equation.
Then the collapse of the wavefunction should be manifested as temporal evolution of the values of the $C_n$'s between $t_i$ and $t_f$, 
probably occurring mostly between $t_a$ (passage of the electron through the slits) and $t_b$ (arrival at the detector screen).
We expect that all but one of the $C_n$'s will vanish at $t=t_f$, but that is a result that should be predicted by the theory rather than imposed as a boundary condition.

But this analysis neglects the entanglement of the electron with the atoms in the detector screen, 
which we posit is essential to predicting the experiment-induced collapse.
For $n=2,3,4, \ldots , N$, let the equilibrium position of atom $n$ be $\vec b_n$, from which we construct the four-vector $b_n = (0,\vec b_n)$, and let its ground and excited states be respectively $\phi(x_n - b_n)$ and $\theta(x_n - b_n)$, and the corresponding energies be $E_0$ and $E_1 > E_0$. 
Then we write the $N$-particle wavefunction as
\begin{multline}
%\begin{split}
\label{two_slit_wavefunction}
\psi(x_1,x_2,x_3,\ldots,x_N)
\\
= \sum_{n=2}^N C_n(x_1^0) \, \xi_n(x_1) \, \eta(x_n - b_n)
\prod^N_{\substack{m=2 \\ m\neq n}}
\phi(x_m - b_m)
\\
= \sum_{n=2}^N C_n(x_1^0) \, \xi_n(x_1)
\prod^N_{m=2} \alpha_{nm}(x_m - b_m)
 \, ,
%\end{split}
\end{multline}
in which $\eta$ is the wavefunction of an atom excited at time $t_b$,
\begin{equation}
\label{define_eta}
\eta(x^0, \vec x) \equiv
\begin{cases}
\phi(x^0, \vec x) & \text{if } x^0 < t_b \\
\theta(x^0, \vec x) & \text{if } x^0 \ge t_b
\end{cases}
\end{equation}
and we define
\begin{equation*}
\alpha_{nm} \equiv \delta_{nm} \eta + (1 - \delta_{nm}) \phi
\end{equation*}
for ease in handling expressions like \eqref{two_slit_wavefunction}.
%(This wavefunction is variational_principle that ignores details of the interaction between the electron and the screen.)
The optimization problem now consists entirely of finding $\{ C_n(t) \}$, since all the other quantities in expression \eqref{two_slit_wavefunction} are known.

Let us suppose that the atomic ground and excited states are orthogonal,
\begin{equation*}
%\label{ground_excited_orthogonality}
\int \rmd^3 x \, \phi^\dagger(t, \vec x) \, \theta(t, \vec x) =
\int \rmd^3 x \, \theta^\dagger(t, \vec x) \, \phi(t, \vec x) =
0 
\quad \forall t
\end{equation*}
and normalized,
\begin{equation}
\label{ground_excited_normalization}
\int \rmd^3 x \, \lvert \phi(t, \vec x) \rvert ^2 =
\int \rmd^3 x \, \lvert \theta(t, \vec x) \rvert ^2 =
1 
\quad \forall t
\end{equation}
We will also take the electron modes to be normalized, and suppose that they are mutually orthogonal. 
The orthogonality is obvious after time $t_a$, because they no longer overlap in space, but for simplicity of this illustrative case we will also suppose that they are orthogonal at earlier times.
Thus
\begin{equation}
\label{electron_normalization}
\int \rmd^3 x \, \xi_m^\dagger(t, \vec x) \, \xi_n(t, \vec x)  =
\delta_{mn}
\quad \forall t, \, \forall m, n \! > \! 1
\, .
\end{equation}

Now the summation in the wavefunction \eqref{two_slit_wavefunction} results in an eightfold sum in $I_K$, but each term in that sum is composed of single-particle factors, thanks to the fact that the operators in
\eqref{define_A_1_general} and \eqref{define_A_2_general} are sums of single-particle operators. With the change of variables $y_{1k}=x_{1k}$ and $y_{mk} = (x_{mk} - b_m) \, \forall m>1$, we find that
\begin{equation*}
I_4(1)=
\mspace{-9mu}
{\sum_{\substack{i_1, i_2, i_3, i_4\\j_1, j_2, j_3, j_4}}}^{\mspace{-18mu} \prime}
D_{i_1 i_2 i_3 i_4 j_1 j_2 j_3 j_4}
\prod_{n=2}^N
E_{n\, i_1 i_2 i_3 i_4 j_1 j_2 j_3 j_4}
\, ,
\end{equation*}
in which we use the prime to signify that the summation variables run from 2 to $N$,
and
\begin{multline}
\label{define_D}
D_{i_1 i_2 i_3 i_4 j_1 j_2 j_3 j_4} \equiv
\int_{\mathcal{R}} 
\rmd^4 y_1 \, \rmd^4 y_2 \, \rmd^4 y_3 \, \rmd^4 y_4
\\
\left[ \prod_{k=1}^4 C_{i_k}^*(y_{k}^0) \, C_{j_k}(y_{k}^0) \, \xi_{i_k}^\dagger(y_{k}) \, \xi_{j_k}(y_{k}) \right]
\\
f( \{y_{q} - y_{r} : 1\! \le \! q \! < \! r \! \le \! 4\} )
\end{multline}
and
\begin{multline*}
%\label{define_E}
%\begin{split}
E_{n\,i_1 i_2 i_3 i_4 j_1 j_2 j_3 j_4}
\equiv
\int_{\mathcal{R}}
\rmd^4 y_1 \, \rmd^4 y_2 \, \rmd^4 y_3 \, \rmd^4 y_4
\\
\left[ \prod_{k=1}^4 \alpha_{{i_k}n}^\dagger(y_{k}) \, \alpha_{{j_k}n}(y_{k}) \right]
f( \{y_{q} - y_{r} : 1\! \le \! q \! < \! r \! \le \! 4\} )
\, .
%\end{split}
\end{multline*}

To simplify further, we note that for any $n$, these integrals have negligible contributions from coordinate values $y_{k}, y_{l} \, (k \ne l)$ in different regions, $y_{k} \in \mathcal{R}_p, y_{l} \in \mathcal{R}_q, p \ne q$.
This is because the spacelike separation constraint (enforced by the factor $f$ in the integrand) restricts the communication between two such points to a temporal separation of the order of the spatial width $\delta x$ of the particle wavefunctions, which is much less than $t_b - t_a$.
Therefore each of these integrals equals the sum of integrals over the subregions $\mathcal{R}_1$, $\mathcal{R}_2$, and $\mathcal{R}_3$.

We further observe that to good approximation, within each of the three subregions the wavefunction product in the integrand in
$E_{n\, i_1 i_2 i_3 i_4 j_1 j_2 j_3 j_4}$ is constant in time, 
%so we can replace the temporal components of the integration variables by $y_{n1}^0$. 
so we can replace $\alpha_{{i_k}n}^\dagger(y_{nk}) \, \alpha_{{j_k}n}(y_{nk})$ 
by $\alpha_{{i_k}n}^\dagger(y_{n1}^0, \vec y_{nk}) \, \alpha_{{j_k}n}(y_{n1}^0, \vec y_{nk})$. 
Then we integrate $f$ on $\rmd y^0_{n2}$, $\rmd y^0_{n3}$, and $\rmd y^0_{n4}$, using \eqref{integrate_final_f_4pt}.
[If $y_1^0$ is within $\pm \delta x$ of $t_i, t_a, t_b,$ or $t_f$, that operation introduces another error of the order of $\delta x$, because limiting the integrals to $\mathcal{R}_1$, $\mathcal{R}_2$, or $\mathcal{R}_3$ denies them positive contributions they would get if their integration range were infinite as in \eqref{integrate_final_f_4pt}.]
Therefore
\begin{equation}
\label{simplify_E}
E_{n\, i_1 i_2 i_3 i_4 j_1 j_2 j_3 j_4}
=
\int_{t_i}^{t_f} \rmd t \,
\prod_{k=1}^4 \int \rmd^3 y \,
\alpha_{{i_k}n}^\dagger(t, \vec y) \, \alpha_{{j_k}n}(t, \vec y)
\, .
\end{equation}

Evaluation of $D_{i_1 i_2 i_3 i_4 j_1 j_2 j_3 j_4}$ is less clear
because the coefficients $C_n$ and the electron modes $\xi_n$ depend on time, 
so the temporal integrations in \eqref{define_D} cannot be done trivially by use of \eqref{integrate_final_f_4pt}. 
However, we expect that the coefficients and the functions $\xi_n$ vary on a timescale of the order of $t_b - t_a$. 
This is much greater than the spatial width of the wavefunction, which is the timescale over which $f$ is nonzero, 
for any choice of positions $\vec y_{11}, \vec y_{12}, \vec y_{13}, \vec y_{14}$ for which the integrand in \eqref{define_D} is not negligible. 
Then we can approximate $D_{i_1 i_2 i_3 i_4 j_1 j_2 j_3 j_4}$ as
\begin{multline*}
D_{i_1 i_2 i_3 i_4 j_1 j_2 j_3 j_4} =
\int_{\mathcal{R}} 
\rmd^4 y_1 \, \rmd^4 y_2 \, \rmd^4 y_3 \, \rmd^4 y_4
\\
\left[ \prod_{k=1}^4 C_{i_k}^*(y_1^0) \, C_{j_k}(y_1^0) \,
\xi_{i_k}^\dagger(y_1^0, \vec y_{k}) \, \xi_{j_k}(y_1^0, \vec y_{k}) \right]
\\
f( \{y_{q} - y_{r} : 1\! \le \! q \! < \! r \! \le \! 4\} )
\end{multline*}
and integrate $f$ on $dy^0_{n2}$, $dy^0_{n3}$, and $dy^0_{n4}$ as before, using \eqref{electron_normalization} to get
\begin{equation}
\label{evaluate_D}
D_{i_1 i_2 i_3 i_4 j_1 j_2 j_3 j_4}
=
\int_{t_i}^{t_f} \rmd t \,
\prod_{k=1}^4 \, \lvert C_{i_k}(t) \rvert ^2
\delta_{i_k j_k}
\, .
\end{equation}
Then $I_4(1)$ becomes
\begin{equation}
\label{evaluate_B}
I_4(1) =
L^{N-1} \int_{t_i}^{t_f} \! \rmd t \, [ \Lambda(t) ] ^4
\end{equation}
in which we define
\begin{equation*}
\Lambda(t) \equiv {\sum_n}^\prime \lvert C_n(t) \rvert ^2
\end{equation*}
and take $L_p$ equal to the duration of $\mathcal{R}_p$:
\begin{equation*}
L_p \equiv
\begin{cases}
t_a - t_i & \text{if } p=1 \\
t_b - t_a & \text{if } p=2 \\
t_f - t_b & \text{if } p=3
\end{cases}
\end{equation*}
and 
\begin{equation*}
L \equiv L_1 + L_2 + L_3 = t_f - t_i
\, .
\end{equation*}

If we analyze $I_4\left[ \left( \delta x^2 \delta p^2 \right)_n \right]$ as we did $I_4(1)$, we find for the $n=1$ term
\begin{multline}
\label{factor_T_1}
I_4\left[ \left( \delta x^2 \delta p^2 \right)_1 \right]
=
\\
\mspace{-9mu}
{\sum_{\substack{i_1, i_2, i_3, i_4\\j_1, j_2, j_3, j_4}}}^{\mspace{-18mu} \prime}
%\mspace{-18mu}
%{\sum_{i_1, i_2, i_3, i_4, j_1, j_2, j_3, j_4}}^{\mspace{-60mu} \prime}
F_{i_1 i_2 i_3 i_4 j_1 j_2 j_3 j_4}
\prod_{n=2}^N
E_{n\, i_1 i_2 i_3 i_4 j_1 j_2 j_3 j_4}
\, ,
\end{multline}
in which we define
\begin{multline*}
%\label{define_F}
F_{i_1 i_2 i_3 i_4 j_1 j_2 j_3 j_4} \equiv
\int_{\mathcal{R}} \rmd^4 y_1 \, \rmd^4 y_2 \, \rmd^4 y_3 \, \rmd^4 y_4 \,
%\\
C_{i_1}^*(y^0_1)
\\
C_{i_2}^*(y^0_2) \, C_{i_3}^*(y^0_3) \, C_{i_4}^*(y^0_4) \, 
C_{j_1}(y^0_1) \, C_{j_2}(y^0_2) \, C_{j_3}(y^0_3) \, C_{j_4}(y^0_4) \\
\left[ \xi_{i_1}^\dagger(y_1) \, \xi_{i_2}^\dagger(y_2) \, 
(y_1^\mu - y_2^\mu) \,(y_1^\nu - y_2^\nu) \, 
\xi_{j_1}(y_1) \, \xi_{j_2}(y_2) \right]
\\
\Big\{ \xi_{i_3}^\dagger(y_3) \, \xi_{i_4}^\dagger(y_4) \, 
[p_{3 \mu}(y_3) - p_{4 \mu}(y_4)] \,
%\\
[p_{3 \nu}(y_3) - p_{4 \nu}(y_4)] \, 
\\
\xi_{j_3}(y_3) \, \xi_{j_4}(y_4) \Big\}
%\\
f( \{y_{q} - y_{r }: 1\! \le \! q \! < \! r \! \le \! 4\} )
\, ,
\end{multline*}
and for the $n>1$ terms the form
\begin{multline}
\label{factor_T_n}
%\begin{split}
I_4\left[ \left( \delta x^2 \delta p^2 \right)_n \right]
=
%&
\mspace{-9mu}
{\sum_{\substack{i_1, i_2, i_3, i_4\\j_1, j_2, j_3, j_4}}}^{\mspace{-18mu} \prime}
%{\sum_{i_1, i_2, i_3, i_4, j_1, j_2, j_3, j_4}}^{\mspace{-60mu}\prime}
D_{i_1 i_2 i_3 i_4 j_1 j_2 j_3 j_4} \,
G_{n\,i_1 i_2 i_3 i_4 j_1 j_2 j_3 j_4}
\\
%&
\qquad \prod_{\substack{p=2 \\ p\neq n}}^N
E_{p\,i_1 i_2 i_3 i_4 j_1 j_2 j_3 j_4}
\, ,
%\end{split}
\end{multline}
where
\begin{multline*}
%\label{define_G}
%\begin{split}
%&
G_{n\,i_1 i_2 i_3 i_4 j_1 j_2 j_3 j_4}
\equiv
\int_{\mathcal{R}} \rmd^4 y_1 \, \rmd^4 y_2 \, \rmd^4 y_3 \, \rmd^4 y_4
\\
%\left[
\alpha_{{i_1}n}^\dagger(y_1) \, \alpha_{{i_2}n}^\dagger(y_2) \,
(y_1^\mu - y_2^\mu) \,(y_1^\nu - y_2^\nu) \,
\alpha_{{j_1}n}(y_1) \, \alpha_{{j_2}n}(y_2)
%\right]
\\
%&
\shoveleft{
%\Big\{
\alpha_{{i_3}n}^\dagger(y_3) \, \alpha_{{i_4}n}^\dagger(y_4) \,
[p_{3 \mu}(y_3) - p_{4 \mu}(y_4)] \,
[p_{3 \nu}(y_3) - p_{4 \nu}(y_4)]
}
\\
\alpha_{{j_3}n}(y_3) \, \alpha_{{j_4}n}(y_4)
\,
%\Big\}
%\\
%&
f( \{y_{q} - y_{r} : 1\! \le \! q \! < \! r \! \le \! 4\} )
\, .
%\end{split}
\end{multline*}
%

%We will now suppose that the $i_k \ne j_k$
%terms in $F_{i_1 i_2 i_3 i_4 j_1 j_2 j_3 j_4}$ are negligible:
We shall see that the principal contributions to $F_{i_1 i_2 i_3 i_4 j_1 j_2 j_3 j_4}$ are from $t>t_a$,
during which time different electronic
wavefunctions $\xi_i, \xi_j$ will be spatially separated.
Then to good approximation
\begin{equation}
\label{F_no_i_ne_j}
F_{i_1 i_2 i_3 i_4 j_1 j_2 j_3 j_4}
=
F_{i_1 i_2 i_3 i_4 i_1 i_2 i_3 i_4}
%\prod_{k=1}^4 \, 
%\delta_{i_k j_k}
\, \delta_{i_1 j_1} \, \delta_{i_2 j_2} \, \delta_{i_3 j_3} \, \delta_{i_4 j_4}
\, .
\end{equation}
%
%The approximation is clearly justified after $t=t_a$ due to the spatial separation of different electronic
%wavefunctions $\xi_i, \xi_j$.
%Before that time it is less accurate,
%but that is acceptable because we will see
%that the principal contributions to $F_{i_1 i_2 i_3 i_4 j_1 j_2 j_3 j_4}$ are from $t>t_a$.

Then due to the delta functions in \eqref{evaluate_D} and \eqref{F_no_i_ne_j}, we will only need to evaluate
$E_{n\,i_1 i_2 i_3 i_4 j_1 j_2 j_3 j_4}$
and $G_{n\,i_1 i_2 i_3 i_4 j_1 j_2 j_3 j_4}$
for the case $i_k = j_k (k=1,2,3,4)$.
We see 
from \eqref{simplify_E}
and the normalization relations \eqref{ground_excited_normalization}
that
\begin{equation}
\label{evaluate_E}
E_{n\,i_1 i_2 i_3 i_4 i_1 i_2 i_3 i_4}
=
L
\, .
\end{equation}

$G_{n\,i_1 i_2 i_3 i_4 i_1 i_2 i_3 i_4}$
is just the four-point single-particle expectation of the single-atom
operator $\left( \delta x^2 \delta p^2 \right)_n$.
Consider first the spatial ($\mu, \nu > 0$) terms in $G_{n\,i_1 i_2 i_3 i_4 i_1 i_2 i_3 i_4}$,
which we will designate as $G^{\,\mathrm{s}}_{n\,i_1 i_2 i_3 i_4}$.
As before, we perform the integrations on $y^0_2, y^0_3,$ and $ y^0_4$ by \eqref{integrate_final_f_4pt},
whereupon the result factors:
\begin{equation*}
%\label{simplify_Gbar1}
G^{\,\mathrm{s}}_{nijkl}
=
\sum^3_{q=1} \sum^3_{r=1}
\int_{t_i}^{t_f} \! \rmd t \,
S^{\,qr}_{nij}(t) \, U^{qr}_{nkl}(t)
\, ,
\end{equation*}
where
\begin{multline*}
%\label{define_S}
S^{\,qr}_{nij}(t) \equiv
\int_{\mathcal{R}} \rmd^3 y \, \rmd^3 z \, 
\alpha_{in}^\dagger(t, \vec y) \, 
\alpha_{jn}^\dagger(t, \vec z) \,
(y_q - z_q)
\\
(y_r - z_r) \,
\alpha_{in}(t, \vec y) \,
\alpha_{jn}(t, \vec z)
%\, ,
\end{multline*}
and
\begin{multline*}
%\label{define_U}
U^{qr}_{nkl}(t) \equiv
\int_{\mathcal{R}} \rmd^3 y \, \rmd^3 z \,
\alpha_{kn}^\dagger(t, \vec y) \, 
\alpha_{ln}^\dagger(t, \vec z) \,
\\
[p_{yq}(t, \vec y) - p_{zq}(t, \vec z)] \,
[p_{yr}(t, \vec y) - p_{zr}(t, \vec z)]
\\
\alpha_{kn}(t, \vec y) \,
\alpha_{ln}(t, \vec z)
\, .
\end{multline*}
$S^{\,qr}_{nij}$ and $U^{qr}_{nkl}$ are components of, respectively,
the position and momentum uncertainties of the wavefunction of atom $n$.
We define the quantities
\begin{multline}
\label{define_delta_x2_phi}
\delta x^2_\phi \equiv
\frac{1}{2}
\int_{\mathcal{R}} \rmd^3 y \, \rmd^3 z \, \phi^\dagger(t, \vec y) \, 
\phi^\dagger(t, \vec z) \,
(y_q - z_q)
\\
(y_q - z_q) \,
\phi(t, \vec y) \,
\phi(t, \vec z)
\, ,
\end{multline}
\begin{multline*}
\delta x^2_\theta \equiv
\frac{1}{2}
\int_{\mathcal{R}} \rmd^3 y \, \rmd^3 z \, \theta^\dagger(t, \vec y) \, 
\theta^\dagger(t, \vec z) \,
(y_q - z_q)
\\
(y_q - z_q) \,
\theta(t, \vec y) \,
\theta(t, \vec z)
\, ,
\end{multline*}
and
\begin{multline*}
%\label{define_delta_x2_phi_theta}
\delta x^2_{\phi \theta} \equiv
\frac{1}{2}
\int_{\mathcal{R}} \rmd^3 y \, \rmd^3 z \, \phi^\dagger(t, \vec y) \, 
\theta^\dagger(t, \vec z) \,
(y_q - z_q)
\\
(y_q - z_q) \,
\phi(t, \vec y) \,
\theta(t, \vec z)
\, ,
\end{multline*}
and analogous quantities $\delta p^2_\phi, \delta p^2_\theta,$ and $\delta p^2_{\phi \theta}$,
none of which we expect will depend on either $t$ or $q \in \{1,2,3\}$.
[Here the factors of $\frac{1}{2}$ arise because each of these integrals
is twice the usual definition of $\delta x^2$:
\begin{equation*}
\begin{split}
{\expectation{\lvert \vec x_1 - \vec x_2 \rvert ^2}}_2
&
= {\expectation{\lvert \vec x_1 \rvert ^2}}_1 - 2 {\expectation{\vec x_1 \cdot \vec x_2}}_2 + {\expectation{\lvert \vec x_2 \rvert ^2}}_1
\\&
= {\expectation{\lvert \vec x_1 \rvert ^2}}_1 - 2 {\expectation{\vec x_1}}_1 \! \cdot \! {\expectation{\vec x_2}}_1 + {\expectation{\lvert \vec x_2 \rvert ^2}}_1
\\&
= 2 \left[ {\expectation{\lvert \vec x_1 \rvert ^2}}_1 - ({\expectation{\vec x_1}}_1)^2 \right]
\\&
= 2 \, {\expectation{\lvert \vec x_1 - {\expectation{\vec x_1}}_1 \rvert ^2}}_1
\, ,
\end{split}
\end{equation*}
for any reasonably defined averages
$\expectationfit{\,}_1$
and
$\expectationfit{\,}_2$ of functions of (respectively) one and two variables.]

Then we expect that for $t_i \le t \le t_b$,
\begin{equation*}
S^{\,qr}_{nij}(t) =
2 \delta_{qr} \, \delta x^2_\phi
%\end{equation*}
%%
%and
%%
%\begin{equation*}
\qquad
U^{qr}_{nkl}(t) =
2 \delta_{qr} \, \delta p^2_\phi
\, ,
\end{equation*}
and for $t_b < t$,
\begin{multline*}
S^{\,qr}_{nij}(t) =
2 \delta_{qr} 
\big\{
(1 - \delta_{ni}) \, (1 - \delta_{nj}) \, \delta x^2_\phi
+
\delta_{ni} \, \delta_{nj} \, \delta x^2_\theta
\\
+
\left[ \delta_{ni} \, (1 - \delta_{nj}) \, + (1 - \delta_{ni}) \, \delta_{nj} \right] \, \delta x^2_{\phi \theta}
\big\}
\end{multline*}
and
\begin{multline*}
U^{qr}_{nkl}(t) =
2 \delta_{qr} 
\big\{
(1 - \delta_{nk}) \, (1 - \delta_{nl}) \, \delta p^2_\phi
+
\delta_{nk} \, \delta_{nl} \, \delta p^2_\theta
\\
+
\left[ \delta_{nk} \, (1 - \delta_{nl}) \, + (1 - \delta_{nk}) \, \delta_{nl} \right] \, \delta p^2_{\phi \theta}
\big\}
\, .
\end{multline*}
Therefore
\begin{multline*}
G^{\,\mathrm{s}}_{nijkl}
=
12 (L_1 + L_2) \, \delta x^2_\phi \, \delta p^2_\phi
\\
\shoveleft{
+ \,
12 L_3 
\big\{
(1 - \delta_{ni}) \, (1 - \delta_{nj}) \, \delta x^2_\phi
+
\delta_{ni} \, \delta_{nj} \, \delta x^2_\theta
}
\\
\shoveright{
+
\left[ \delta_{ni} \, (1 - \delta_{nj}) \, + (1 - \delta_{ni}) \, \delta_{nj} \right] \, \delta x^2_{\phi \theta}
\big\}
}
\\
\shoveleft{
\quad \big\{
(1 - \delta_{nk}) \, (1 - \delta_{nl}) \, \delta p^2_\phi
+
\delta_{nk} \, \delta_{nl} \, \delta p^2_\theta
}
\\
+
\left[ \delta_{nk} \, (1 - \delta_{nl}) \, + (1 - \delta_{nk}) \, \delta_{nl} \right] \, \delta p^2_{\phi \theta}
\big\}
\end{multline*}
and
\begin{equation}
\begin{split}
\label{sum_G1}
{\sum_n}^\prime
G^{\,\mathrm{s}}_{nijkl}
&
=
12 (N-1) \,
L \,
\delta x^2_\phi \, \delta p^2_\phi 
+ \Or(L_3 \delta x^2_\phi \, \delta p^2_\phi)
\\
&
\simeq
3N
L 
+ \Or(L)
\, ,
\end{split}
\end{equation}
supposing that the ground and excited states $\phi$ and $\theta$ are near minimal-uncertainty states for the atoms.

Now
\begin{multline*}
%\label{combine_F}
F_{i_1 i_2 i_3 i_4 i_1 i_2 i_3 i_4}
=
\int_{\mathcal{R}} \rmd^4 y_1 \, \rmd^4 y_2 \, \rmd^4 y_3 \, \rmd^4 y_4 \,
\\
\lvert C_{i_1}(y^0_1) \rvert^2 \, \lvert C_{i_2}(y^0_2) \rvert^2 \, \lvert C_{i_3}(y^0_3) \rvert^2 \, \lvert C_{i_4}(y^0_4) \rvert^2
\\
%\left[
\xi_{i_1}^\dagger(y_1) \, \xi_{i_2}^\dagger(y_2) \, 
(y_1^\mu - y_2^\mu) \,(y_1^\nu - y_2^\nu) \, 
\xi_{i_1}(y_1) \, \xi_{i_2}(y_2)
%\right]
\\
\shoveleft{
\quad
%\Big\{
\xi_{i_3}^\dagger(y_3) \, \xi_{i_4}^\dagger(y_4) \, 
[p_{3 \mu}(y_3) - p_{4 \mu}(y_4)]
}
\\
\shoveright{
[p_{3 \nu}(y_3) - p_{4 \nu}(y_4)] \, 
\xi_{i_3}(y_3) \, \xi_{i_4}(y_4)
%\Big\}
}
\\
f( \{y_{q} - y_{r }: 1\! \le \! q \! < \! r \! \le \! 4\} )
\, .
\end{multline*}
We will designate the $\mu, \nu > 0$ terms in $F_{i_1 i_2 i_3 i_4 i_1 i_2 i_3 i_4}$
as $F^{\,\mathrm{s}}_{i_1 i_2 i_3 i_4}$.
We note that in those terms, the only
time-dependent factors in the integrand besides $f$ are of the form $\lvert C_{i_k} \rvert^2$ and 
$\xi_{i_k}^\dagger \xi_{i_k}$, both of which vary much more slowly than $f$.
Then we can approximate the time coordinate of those factors by $y^0_1$; that is,
for $k=2,3,4$ we replace $y_k = (y_k^0, \vec y_k)$ by $(y^0_1, \vec y_k)$
except within the arguments of $f$. This allows us to integrate over the
temporal variables $y^0_2, y^0_3,$ and $ y^0_4$ as before, with the result that 
\begin{equation*}
%\label{simplify_Fbar1}
F^{\,\mathrm{s}}_{ijkl}
=
\sum^3_{q=1} \sum^3_{r=1}
\int_{t_i}^{t_f} \! \rmd t 
\left[
\prod^4_{k=1}
\lvert C_{i_k}(t) \rvert ^2
\right]
V^{qr}_{ij}(t) \, W^{qr}_{kl}(t)
\, ,
\end{equation*}
where
\begin{multline*}
%\label{define_V}
V^{qr}_{ij}(t) \equiv
\int_{\mathcal{R}} \rmd^3 y \, \rmd^3 z \, 
\xi_{i}^\dagger(t, \vec y) \, 
\xi_{j}^\dagger(t, \vec z) \,
(y_q - z_q)
\\
(y_r - z_r) \,
\xi_{i}(t, \vec y) \,
\xi_{j}(t, \vec z)
\end{multline*}
and
\begin{multline*}
%\label{define_W}
W^{qr}_{kl}(t) \equiv
\int_{\mathcal{R}} \rmd^3 y \, \rmd^3 z \,
\xi_k^\dagger(t, \vec y) \, 
\xi_l^\dagger(t, \vec z) \,
[p_{yq}(\vec y) - p_{zq}(t, \vec z)]
\\
[p_{yr}(\vec y) - p_{zr}(t, \vec z)] \,
\xi_k(t, \vec y) \,
\xi_l(t, \vec z)
\, .
\end{multline*}
By symmetry, these expressions vanish whenever $q \ne r$;
otherwise, they are twice the squared position and momentum uncertainties of the electron between states
$i$ and $j$ at time $t$
[compare \eqref{define_delta_x2_phi}].
Then we expect that the position-momentum uncertainty in the $x$ direction will take the minimum value:
\begin{equation*}
V^{11}_{ij}(t) \, W^{11}_{kl}(t)
\simeq
2^2 
\left( \frac{1}{4} \right)
= 1
\, .
\end{equation*}
If $q=2$ or $3$, we add to that uncertainty a macroscopic term that we can estimate from simple problem geometry:
\begin{multline*}
V^{qq}_{ij}(t) \, W^{qq}_{kl}(t)
\simeq
\\
\begin{cases}
1 & \text{if } t < t_a \\
m_e^2 (b_{iq} - b_{jq})^2 (b_{kq} - b_{lq})^2 \frac{(t - t_a)^2}{(t_b - t_a)^4} + 1 & \text{if } t_a < t < t_b \\
2 (b_{iq} - b_{jq})^2 \, \delta p^2_\xi + 1 & \text{if } t_b < t
\, ,
\end{cases}
\end{multline*}
where $\delta p^2_\xi$ is the momentum uncertainty (in the $y$ or $z$ direction)
of the electron after it interacts with the detector screen.
Therefore
\begin{multline}
\label{evaluate_Fbar1}
F^{\,\mathrm{s}}_{i_1 i_2 i_3 i_4}
=
3
\int_{t_i}^{t_f} \! \rmd t \,
\prod^4_{k=1} \,
\lvert C_{i_k}(t) \rvert ^2
\\
\shoveleft{
+
m_e^2 \,
\left[
\sum_{q=2}^3
(b_{i_1 q} - b_{i_2 q}) ^2
(b_{i_3 q} - b_{i_4 q}) ^2
\right]
}
\\
\shoveright{
\int_{t_a}^{t_b} \! \rmd t \,
\frac{
(t - t_a)^2
}
{
(t_b - t_a)^4
}
\prod^4_{k=1} \,
\lvert C_{i_k}(t) \rvert ^2
}
\\
+
2 \, \delta p_\xi^2
\left[
\sum_{q=2}^3
(b_{i_1 q} - b_{i_2 q}) ^2
\right]
\int_{t_b}^{t_f} \! \rmd t \,
\prod^4_{k=1} \,
\lvert C_{i_k}(t) \rvert ^2
\, .
\end{multline}

As to the temporal ($\mu = \nu = 0$) terms in $G_{n\,i_1 i_2 i_3 i_4 i_1 i_2 i_3 i_4}$,
which we will designate as $G^{\,\mathrm{t}}_{n\,i_1 i_2 i_3 i_4}$
(the terms with $\mu = 0, \nu \ne 0$ and with $\mu \ne 0, \nu= 0$ vanish by symmetry),
we cannot perform the time integrations as readily as before because of the time-dependent expressions 
$(y_1^0 - y_2^0)$ and $(y_3^0 - y_4^0)$ in the integrand. However, we know that the factor 
$f(y_1-y_2, \ldots y_3-y_4)$
is zero unless $\lvert y_1^0 - y_2^0 \rvert$ does not exceed $\lvert \vec y_1 - \vec y_2 \rvert$,
which is limited to values of the order of $\sqrt{\delta x^2_\phi}$ wherever the integrand is nonzero.
Therefore we can estimate
\begin{multline*}
G^{\,\mathrm{t}}_{nijkl}
\simeq 
\left( 2 \, \delta x^2_\phi \right)
\int_{\mathcal{R}} \rmd^4 y_1 \, \rmd^4 y_2 \, \rmd^4 y_3 \, \rmd^4 y_4
\\
\shoveleft{
\alpha_{in}^\dagger(y_1) \, \alpha_{jn}^\dagger(y_2) \,
\alpha_{in}(y_1) \, \alpha_{jn}(y_2) \,
\alpha_{kn}^\dagger(y_3) \, \alpha_{ln}^\dagger(y_4)
}
\\
[p_{30}(y_3) - p_{40}(y_4)]^2 \,
\alpha_{kn}(y_3) \, \alpha_{ln}(y_4)
\\
\shoveright{
f( \{y_{q} - y_{r} : 1\! \le \! q \! < \! r \! \le \! 4\} )
}
\\
\shoveleft{
=
2 \, \delta x^2_\phi
\int_{t_i}^{t_f} \! \rmd t \, \int_{\mathcal{R}} \rmd^3 y_3 \, \rmd^3 y_4 \,
\alpha_{kn}^\dagger(t, \vec y_3) \, \alpha_{ln}^\dagger(t, \vec y_4)
}
\\
[p_{30}(t, \vec y_3) - p_{40}(t, \vec y_4)]^2 \,
\alpha_{kn}(t, \vec y_3) \, \alpha_{ln}(t, \vec y_4)
\\
\shoveleft{
=
4 \, \delta x^2_\phi
(L_1 + L_2) \,
\delta E^2_\phi
}
\\
+
4 \, \delta x^2_\phi \,
L_3
\big\{
(1 - \delta_{nk}) \, (1 - \delta_{nl}) \, \delta E^2_\phi
+
\delta_{nk} \, \delta_{nl} \, \delta E^2_\theta
\\
+
\left[ \delta_{nk} \, (1 - \delta_{nl}) \, + (1 - \delta_{nk}) \, \delta_{nl} \right] \, \delta E^2_{\phi \theta}
\big\}
\end{multline*}
in which $\delta E^2_\phi, \delta E^2_\theta,$ and $\delta E^2_{\phi \theta}$
are defined just as $\delta p^2_\phi, \delta p^2_\theta,$ and $\delta p^2_{\phi \theta}$
but for the $\mu=0$ component of the energy-momentum four-vector $p^\mu$.

We expect that
\begin{equation*}
\delta x^2_\phi \, \delta E^2_\phi = \delta x^2_\phi \, \delta E^2_\theta = \Or(1)
\, ,
\end{equation*}
but
\begin{equation*}
\delta E^2_{\phi \theta} = (E_1 - E_0)^2
\, .
\end{equation*}
Then we see that
\begin{multline*}
{\sum_n}^\prime
G^{\,\mathrm{t}}_{nijkl}
\simeq
4 \, \delta x^2_\phi
\big[
(N-1) \, L \,
\delta E^2_\phi
\\
+
2(1 - \delta_{kl}) L_3 \, 
(E_1 - E_0)^2
\big]
+ \Or(L_3)
\, .
\end{multline*}
Combining this with the $\mu, \nu \ne 0$ terms from \eqref{sum_G1},
\begin{multline}
\label{sum_G}
{\sum_n}^\prime
G_{n\,i_1 i_2 i_3 i_4 i_1 i_2 i_3 i_4}
\simeq 
\left(
3 +
4 \, \delta x^2_\phi \, \delta E^2_\phi
\right)
NL
\\
+
8 (1 - \delta_{i_3 i_4}) \,
\delta x^2_\phi L_3 (E_1 - E_0)^2
+ \Or(L)
\, .
\end{multline}

By similar reasoning, we can approximate the $\mu = \nu = 0$ terms in 
$F_{i_1 i_2 i_3 i_4 i_1 i_2 i_3 i_4}$ as
$F^{\,\mathrm{t}}_{i_1 i_2 i_3 i_4}$, defined by
\begin{multline*}
F^{\,\mathrm{t}}_{i j k l}
\simeq
\int_{\mathcal{R}} \rmd^4 y_1 \, \rmd^4 y_2 \, \rmd^4 y_3 \, \rmd^4 y_4 \,
\\
\lvert C_{i}(y^0_1) \rvert^2 \, \lvert C_{j}(y^0_1) \rvert^2 \, 
\lvert C_{k}(y^0_1) \rvert^2 \, \lvert C_{l}(y^0_1) \rvert^2 \,
V^{11}_{i j}(y^0_1)
\\
\xi_{i}^\dagger(y_1) \, \xi_{j}^\dagger(y_1^0, \vec y_2) \, 
\xi_{i}(y_1) \, \xi_{j}(y_1^0, \vec y_2)
\\
\xi_{k}^\dagger(y_1^0, \vec y_3) \, \xi_{l}^\dagger(y_1^0, \vec y_4) \, 
[p_3^0(y_1^0, \vec y_3) - p_4^0(y_1^0, \vec y_4)]^2
\\
\xi_{k}(y_1^0, \vec y_3) \, \xi_{l}(y_1^0, \vec y_4) \,
f( \{y_{q} - y_{r }: 1\! \le \! q \! < \! r \! \le \! 4\} )
\\
\shoveleft{
\qquad =
\int_{t_i}^{t_f} \rmd t \int_{\mathcal{R}} \rmd^3 y_1 \, \rmd^3 y_2 \, \rmd^3 y_3 \, \rmd^3 y_4
}
\\
\lvert C_{i}(t) \rvert^2 \, \lvert C_{j}(t) \rvert^2 \, 
\lvert C_{k}(t) \rvert^2 \, \lvert C_{l}(t) \rvert^2 \,
V^{11}_{i j}(t)
 \\
\xi_{i}^\dagger(t, \vec y_1) \, \xi_{j}^\dagger(t, \vec y_2) \, 
\xi_{i}(t, \vec y_1) \, \xi_{j}(t, \vec y_2)
\\
\xi_{k}^\dagger(t, \vec y_3) \, \xi_{l}^\dagger(t, \vec y_4) \, 
[p_3^0(t, \vec y_3) - p_4^0(t, \vec y_4)]^2
\\
\xi_{k}(t, \vec y_3) \, \xi_{l}(t, \vec y_4)
\\
%\shoveleft{
%\qquad =
%\int_{t_i}^{t_f} \rmd t \int_{\mathcal{R}} \rmd^3 y_3 \, \rmd^3 y_4
%}
%\\
%\lvert C_{i}(t) \rvert^2 \, \lvert C_{j}(t) \rvert^2 \, 
%\lvert C_{k}(t) \rvert^2 \, \lvert C_{l}(t) \rvert^2 \,
%V^{11}_{i j}(t)
% \\
%\xi_{k}^\dagger(t, \vec y_3) \, \xi_{l}^\dagger(t, \vec y_4) \, 
%[p_3^0(t, \vec y_3) - p_4^0(t, \vec y_4)]^2
%\\
%\xi_{k}(t, \vec y_3) \, \xi_{l}(t, \vec y_4)
%\\
\shoveleft{
\qquad =
2 \, \delta E^2_\xi
}
\\
\int_{t_i}^{t_f} \rmd t \,
%\\
\lvert C_{i}(t) \rvert^2 \, \lvert C_{j}(t) \rvert^2 \, 
\lvert C_{k}(t) \rvert^2 \, \lvert C_{l}(t) \rvert^2 \,
%\\
V^{11}_{i j}(t) \,
 \, .
\end{multline*}
Since the electron is nonrelativistic, 
%
%\begin{equation*}
$V^{11}_{i j} \ll (t_f- t_i)^2$,
%\, ,
%\end{equation*}
%
which must be less than $2 \, \delta t^2_\xi$, (a constant times the squared lifetime of the electron).
Since for a minimum-uncertainty state
\begin{equation*}
\delta t^2_\xi \, \delta E^2_\xi
\simeq
\frac{1}{4}
\, ,
\end{equation*}
we conclude that
\begin{equation*}
F^{\,\mathrm{t}}_{i_1 i_2 i_3 i_4}
\ll
\frac{1}{2}
\int_{t_i}^{t_f} \rmd t 
\prod^4_{k=1} \,
\lvert C_{i_k}(t) \rvert ^2
\, .
\end{equation*}
Then to good approximation, $F_{i_1 i_2 i_3 i_4 i_1 i_2 i_3 i_4}$ is equal to the expression
on the RHS of \eqref{evaluate_Fbar1}.

Now we use \eqref{F_no_i_ne_j} and \eqref{evaluate_E} to simplify \eqref{factor_T_1},
and substitute into it expression \eqref{evaluate_Fbar1} for $F_{i_1 i_2 i_3 i_4 i_1 i_2 i_3 i_4}$.
The result is
\begin{multline*}
%\label{evaluate_T_1}
I_4\left[ \left( \delta x^2 \delta p^2 \right)_1 \right]
=
L^{N-1}
{\sum_{i_1, i_2, i_3, i_4}}^{\mspace{-18mu} \prime}
\mspace{12mu}
\Bigg\{
3
\int_{t_i}^{t_f} \! \rmd t \,
\prod^4_{k=1} \,
\lvert C_{i_k}(t) \rvert ^2
\\
\shoveleft{
+
m_e^2 \,
\left[
\sum_{q=2}^3
(b_{i_1 q} - b_{i_2 q}) ^2
(b_{i_3 q} - b_{i_4 q}) ^2
\right]
}
\\
\shoveright{
\int_{t_a}^{t_b} \! \rmd t \,
\frac{
(t - t_a)^2
}
{
(t_b - t_a)^4
}
\prod^4_{k=1} \,
\lvert C_{i_k}(t) \rvert ^2
}
\\
+
2 \, \delta p_\xi^2
\left[
\sum_{q=2}^3
(b_{i_1 q} - b_{i_2 q}) ^2
\right]
\int_{t_b}^{t_f} \! \rmd t \,
\prod^4_{k=1} \,
\lvert C_{i_k}(t) \rvert ^2
\Bigg\}
\, .
\end{multline*}
We also sum \eqref{factor_T_n} on $n$ and substitute into it expressions
\eqref{evaluate_D}, \eqref{evaluate_E} and \eqref{sum_G}:
\begin{multline*}
%\label{sum_T_n}
{\sum_n}^{\, \prime}
I_4\left[ \left( \delta x^2 \delta p^2 \right)_n \right]
=
L^{N-1}
\mspace{-12mu}
{\sum_{i_1, i_2, i_3, i_4}}^{\mspace{-18mu}\prime}
\mspace{9mu}
\bigg[
\left(
3 +
4 \, \delta x^2_\phi \, \delta E^2_\phi
\right)
N
\\
+
8 (1 - \delta_{i_3 i_4})
\frac
{L_3}
{L}
\, 
\delta x^2_\phi (E_1 - E_0)^2
+ \Or(1)
\bigg]
\\
\int_{t_i}^{t_f} \rmd t \,
\prod_{k=1}^4 \, \lvert C_{i_k}(t) \rvert ^2
\, ,
\end{multline*}

Using these results and expression \eqref{evaluate_B} for $I_4(1)$ in
\eqref{define_<O>_K} and \eqref{define_A_2_general}, we find that
\begin{multline*}
%\label{evaluate_A_2}
A_2 =
\left\{ \int_{t_i}^{t_f} \! \rmd t \, [ \Lambda(t) ] ^4 \right\} ^{-1}
{\sum_{i_1, i_2, i_3, i_4}}^{\mspace{-18mu} \prime}
\mspace{12mu}
\Bigg\{
\Bigg[
\left(
3 +
4 \, \delta x^2_\phi \, \delta E^2_\phi
\right)
N
\\
+
8 (1 - \delta_{i_3 i_4})
\frac
{L_3}
{L}
\, 
\delta x^2_\phi (E_1 - E_0)^2
+ \Or(1)
\Bigg]
\int_{t_i}^{t_f} \! \rmd t \,
\prod^4_{k=1} \,
\lvert C_{i_k}(t) \rvert ^2
\\
\shoveleft{
+
m_e^2 \,
\left[
\sum_{q=2}^3
(b_{i_1 q} - b_{i_2 q}) ^2
(b_{i_3 q} - b_{i_4 q}) ^2
\right]
}
\\
\shoveright{
\int_{t_a}^{t_b} \! \rmd t \,
\frac{
(t - t_a)^2
}
{
(t_b - t_a)^4
}
\prod^4_{k=1} \,
\lvert C_{i_k}(t) \rvert ^2
}
\\
+
2 \, \delta p_\xi^2 \,
\left[
\sum_{q=2}^3
(b_{i_1 q} - b_{i_2 q}) ^2
\right]
\int_{t_b}^{t_f} \! \rmd t \,
\prod^4_{k=1} \,
\lvert C_{i_k}(t) \rvert ^2
\Bigg\}
\, .
\end{multline*}

Recall that our objective is to demonstrate the collapse of the wavefunction given in \eqref{two_slit_wavefunction}
to a single term in that sum, as time advances from $t_i$ to $t_f$. It will therefore be helpful to consider
$w_i(t) \equiv \lvert C_i(t) \rvert^2 / \Lambda(t)$, the weights of modes $i$ relative to the total at any time $t$, 
which are nonnegative real numbers satisfying $\sum^\prime w_j = 1$ at any time $t$.
Then $A_2$ takes the form
\begin{equation}
\label{A_2_a_form}
A_2
=
\int_{t_i}^{t_f} \! \rmd t
{\sum_{i, j, k, l}}^{\prime}
w_i(t) \, w_j(t) \, w_k(t) \, w_l(t) \,
a_{i j k l}(t)
\end{equation}
with
\begin{equation*}
a_{i j k l}(t)
\equiv
\begin{cases}
a^{(1)}_{i j k l} & \text{if } t < t_b \\
a^{(1)}_{i j k l} + a^{(2)}_{i j k l}(t) & \text{if } t_b < t < t_a \\
a^{(1)}_{i j k l} + a^{(3)}_{i j k l} & \text{if } t_a < t
\, ,
\end{cases}
\end{equation*}
\begin{multline*}
%\label{define_a^1}
a^{(1)}_{i j k l}
\equiv
\left(
3
+
4 \, \delta x^2_\phi \, \delta E^2_\phi
\right)
N
\\
+
8 (1 - \delta_{k l}) 
\frac{L_3}{L}
\delta x^2_\phi (E_1 - E_0)^2
\, ,
\end{multline*}
\begin{equation*}
a^{(2)}_{i j k l}(t)
\equiv
m_e^2 \,
\frac{
(t - t_a)^2
}
{
(t_b - t_a)^4
}
\,
\sum_{q=2}^3
(b_{i q} - b_{j q}) ^2
(b_{k q} - b_{l q}) ^2
\, ,
\end{equation*}
and
\begin{equation*}
a^{(3)}_{i j k l}
\equiv
2 \, \delta p_\xi^2 \,
\sum_{q=2}^3
(b_{i q} - b_{j q}) ^2
\, ,
\end{equation*}
neglecting $\Or(1)$.
$a^{(1)}_{i j k l}$ is our estimate of the atomic contributions to the total wavefunction uncertainty,
with the first term estimating the zero-point motions, and the second term accounting for the
macroscopic energy uncertainty in a superposition of the states in which atom $i, j, k,$ or $l$ is excited.
Terms $a^{(2)}_{i j k l}$ and $a^{(3)}_{i j k l}$ are the estimated position-momentum uncertainty due to
the undetermined electron trajectory; the uncertainty of its zero-point motion was included in the 
$\Or(1)$ terms that were dropped.

Now 
%all the factors in the integrands in \eqref{A_2_a_form} are positive definite.
the VP attempts to choose $\{ w_j(t) \}$ so as to minimize $A_2$.
If that were the only term in the VP, it would allow no more than one of those weights to be nonzero,
so as to avoid contributions to \eqref{A_2_a_form} from 
$a^{(2)}_{i j k l}, a^{(3)}_{i j k l},$ or the second term of $a^{(1)}_{i j k l}$. 
As shown in \cite{Harrison_2012a}, however, the $A_1$ term in the VP penalizes rapid changes
in the wavefunction. Since the initial values of the weights are constrained by
\eqref{decompose_xi_SQM} to describe the SQM diffraction pattern, 
they must evolve continuously from those values toward the solution
\begin{equation*}
w_i(t) =
\begin{cases}
1  & \text{if } i=j \\
0  & \text{if } i \neq j
\end{cases}
\end{equation*}
for some fixed $j$.
Presumably that evolution is complete, or approximately so, by $t_f$.
If we had a precisely defined experimental setup
[and an exact form for the function $f_K$ in \eqref{define_I_K}],
we could perform more careful analysis of the VP,
including the $A_1$ term, which would allow us to test that approximation.

But \cite{Harrison_2012a} also shows that $A_1$ will enforce the Born rule;
therefore the initial coefficient values in \eqref{decompose_xi_SQM}
will describe the outcome probabilities.
That means that in many realizations, the measured positions will sum to the interference pattern predicted by the
Dirac equation, in agreement with experiments actually conducted. \cite{Merli_Missiroli_Pozzi,Tonomura}

Now suppose the experiment were carried out with a different type of detector, one that detects the wave structure of an interference pattern produced on the screen without identifying a specific location for the electron.
Hypothetically, we might imagine a microwave cavity or waveguide designed to support a mode with wavelength comparable to that of the expected interference pattern.
Suppose the screen were made of a suitable transducer material that excites the electric field in the cavity in geometrical conformity to the incident wave pattern on the screen.
Then the wavefunctions $\theta$ and $\phi$ in \eqref{define_eta} would correspond to electromagnetic modes of the cavity,
localized not in position, but in wavevector space.
The electron would most simply be described in terms of a basis parametrized by position in the $x^1$ direction and wavenumber in the $x^2$ and $x^3$ directions.
Then the derivation would proceed as it did before, with the result that the electron's wavefunction, at the time of its arrival at the screen, would have a well-defined transverse wavelength rather than a position.

Having analyzed the two-slit experiment, we take the opportunity to consider Wheeler's ``delayed-choice'' variant \cite{Wheeler_1979} of that experiment.
The essential
element is that some aspect of the detection system is modified during the interval between $t_a$ and $t_b$.
This variant is particularly puzzling because the original form of the two-slit experiment suggests, when considered from the standpoint
of everyday experience with classical macroscopic objects, that the electron ``chooses'' whether to pass through one slit
or both based on the nature of the detection system. By delaying the choice of detector system until after the electron passes the
slit-containing barrier, Wheeler's variant challenges that description of the experiment.

The analysis just presented of the two-slit experiment can be applied virtually unchanged to Wheeler's delayed-choice variant.
Clearly the detection apparatus would have to be represented differently than we have done, for times before and during
the modification of that apparatus, in accordance with the definition of the experimental procedure. Nevertheless, the
terms contributing to $A_2$ that actually cause collapse due to their dependence on the modal content of the wavefunction
will depend on the state of the detector near and after $t_b$, and will be just as we have computed them.
The parts of the system-detector wavefunction that must be added to represent
the ``delayed-choice'' features of the experiment will not affect the collapse itself, or the choice into which state the electron
ends up. Therefore Wheeler's experiment has exactly the same outcome as the original version.

Another way to understand this is that Nature minimizes the functional in the VP by considering the entire range of space
\emph{and time} that participate in the experiment; thus the choice is not made \emph{at some instant in time}. 
It is meaningless to ask ``when'' the electron made a choice;
the best description may be that the decision was made \emph{outside of time}.

\section{Summary and conclusions}

In an earlier paper, \cite{Harrison_2012a} this author proposed a variational-principle formulation of quantum mechanics as an alternative to SQM. The theory was intended to encompass both the measurement and non-measurement regimes, traditionally described in SQM by two very different rules. In order to test the VP, we have here applied it to the electron two-slit experiment, which has long been regarded as posing a conceptual challenge to SQM.

We have shown that, subject to our idealization of the experiment and certain approximations made in the calculation,
measurements made employing a position-sensitive detector screen will in fact show that
any single electron is intercepted at a well-defined position.
The accumulation of many such results in repetitions of the experiment will produce the interference pattern described by the SQM wave equation.
Both these predictions agree with the experimental record.
We have argued that with a detector that was truly sensitive to the transverse wavelength of the electron rather than its position, the electron
would collapse to a single state with a well-defined wavelength. We have pointed out that Wheeler's ``delayed-choice'' experiment
\cite{Wheeler_1979} is
predicted to have the same result as the original form of the measurement. We regard this as a manifestation of a process in which the
determination (``choice,'' in anthropomorphic terms) of the experimental outcome is not made at an instant of time, but rather outside of time;
this disposes of the puzzling character of Wheeler's innovation.

Finally, we observe that the solution of the VP resolves the conceptual difficulties described by Feynman in his discussion of the two-slit experiment. \cite{Feynman} He argued that there was no plausible way to describe the electron's path that would be consistent with the
experimental outcomes. But we have a plausible description, which is that the electron wave passes through both slits, satisfying a wave
``equation'' (the VP) that describes the wave converging to the vicinity of a single atom of the detection screen.

Of course, as pointed out in \cite{Harrison_2012a}, the theory exhibits retrocausation, which is a different conceptual challenge!

\begin{acknowledgments}
Los Alamos National Laboratory, an affirmative action/equal opportunity employer, is operated by Los Alamos National Security, LLC, for the National Nuclear Security Administration (NNSA) of the U.S. Department of Energy under contract DE-AC52-06NA25396.
The author appreciates the support of the NNSA Advanced Simulation and Computing (ASC) program; 
helpful discussions with Salman Habib, Robin Blume-Kohout, Terry Goldman, Howard Brandt, Baolian Cheng and David Sigeti; 
review of an earlier draft by Jean-Francois Van Huele; 
and detailed discussions with Dale W. Harrison and B. Kent Harrison over a long period of time. 
He is, however, solely responsible for any errors or deficiencies in the work.
\end{acknowledgments}

\bibliography{harr0417}

\end{document}